\documentclass[10.5pt,superscriptaddress,longbibliography,twocolumn]{revtex4-2}
\usepackage{amsmath,amssymb,amsfonts,float,graphics,epsfig,epstopdf,color,verbatim,tabularx,bm,multirow,appendix}
\usepackage[utf8]{inputenc}
\usepackage[T1]{fontenc}
\usepackage{xcolor}
\usepackage{dsfont}
\usepackage{textcomp}
\usepackage{yfonts}
\usepackage{bm}
\usepackage{subfigure}
\usepackage{mathrsfs}
\usepackage{graphicx}
\usepackage{verbatim}
\usepackage{hyperref}
\usepackage{multirow}
\usepackage{braket}
\usepackage[normalem]{ulem}
\usepackage{tikz}
\usetikzlibrary{calc}
\usetikzlibrary{shapes.multipart}
\usepackage[final,commandnameprefix=ifneeded,defaultcolor=red]{changes}

\renewcommand{\vec}[1]{{\mathbf #1}}

\newcommand{\comments}[1]{}

\def\H{\mathcal{H}}

\newcommand{\stkout}[1]{\ifmmode\text{\sout{\ensuremath{#1}}}\else\sout{#1}\fi}

\makeatletter
\def\l@subsubsection#1#2{}
\makeatother

\begin{document}
	
\title{Finite-temperature critical behaviors in 2D long-range quantum Heisenberg model}
	
	\author{Jiarui Zhao}
	\affiliation{Department of Physics and HKU-UCAS Joint Institute of Theoretical and Computational Physics, The University of Hong Kong, Pokfulam Road, Hong Kong SAR, China}
	
	\author{Menghan Song}
	\affiliation{Department of Physics and HKU-UCAS Joint Institute of Theoretical and Computational Physics, The University of Hong Kong, Pokfulam Road, Hong Kong SAR, China}
		
	\author{Yang Qi}
	\affiliation{State Key Laboratory of Surface Physics, Fudan University, Shanghai 200438, China}
	\affiliation{Center for Field Theory and Particle Physics, Department of Physics, Fudan University, Shanghai 200433, China}
	\affiliation{Collaborative Innovation Center of Advanced Microstructures, Nanjing 210093, China}

 \author{Junchen Rong}
  \email{junchenrong@gmail.com}	
 \affiliation{Institut des Hautes \'Etudes Scientifiques, 91440 Bures-sur-Yvette, France}

	\author{Zi Yang Meng}
	\email{zymeng@hku.hk}
	\affiliation{Department of Physics and HKU-UCAS Joint Institute of Theoretical and Computational Physics, The University of Hong Kong, Pokfulam Road, Hong Kong SAR, China}
	
	\begin{abstract}
The Mermin-Wagner theorem states that spontaneous continuous symmetry breaking is prohibited in systems with short-range interactions at spatial dimension $D\le 2$. For long-range interactions with a power-law form ($1/r^{\alpha}$), the theorem further forbids ferromagnetic or antiferromagnetic order at finite temperature when $\alpha\ge 2D$. However, the situation for $\alpha \in (2,4)$ at $D=2$ is not covered by the theorem. To address this, we conduct large-scale quantum Monte Carlo simulations and field theoretical analysis. Our findings show spontaneous breaking of $SU(2)$ symmetry in the ferromagnetic Heisenberg model with $1/r^{\alpha}$-form long-range interactions at $D=2$. We determine critical exponents through finite-size analysis for $\alpha<3$ (above the upper critical dimension with Gaussian fixed point) and $3\le\alpha<4$ (below the upper critical dimension with non-Gaussian fixed point). These results reveal new critical behaviors in 2D long-range Heisenberg models, encouraging further experimental studies of quantum materials with long-range interactions beyond the Mermin-Wagner theorem's scope. 
\end{abstract}
	\date{\today}
	\maketitle
	
In recent years, the importance of the studies on \replaced{long-range(LR)}{LR} lattice models have been gradually noticed, due to the fact that they exhibit intrinsically different  properties from their short-ranged\added{(SR)} counterparts. For example, LR Heisenberg models at spatial dimension $D=2$ acquires anomalous magnon dispersion different from the linear and quadratic spin-waves in the SR antiferromagnetic and ferromagnetic models~\cite{DiesselGeneralized2022,songDynamical2023}. In addition, the violation of Mermin-Wagner theorem and unconventional critical properties in LR systems also attracted much attention in investigations of both quantum spin models and interacting fermionic models~\cite{CriticalExponents1972,sak1973recursion,aizenman1988critical,lohmann2017critical,sakLow1977,Slade_2017,DefenuLong2021,Eduardo2021,BirnkammerCharacterzing2022,PeterAnomalous2012,zhuQuantum2016,AdelhardtContinuously2022,HamerThird-order1992,jenkins2022breaking,Maghrebi2017,KoziolQuantum2021,Flores-Sola2015,Eduardo2021,wangOn2022,WeberDissipation2022,wernerQuantum2005,laflorencie2005critical}.

These phenomena also have immediate experimental relevance. Due to the fast development in the Rydberg atom arrays~\cite{samajdar2021quantum,yan2022triangular,Semeghini21,Roushan21,yan2022triangular,yanEmergent2023}, the magic angle
twisted bilayer Graphene and other 2D quantum
moir\'e materials~\cite{tramblyLocalization2010,rafiMoire2011,tramblyNumerical2012,rozhkovElectronic2016,caoUnconventional2018,caoCorrelated2018,xieSpectroscopic2019,luSuperconductors2019,liaoValence2019,yankowitzTuning2019,yankowitzTuning2019,tomarkenElectronic2019,caoStrange2020,shenCorrelated2020,kevinStrongly2020,chatterjeeSkyrmion2022,khalafSoftmodes2020,xieNature2020,rozenEntropic2021,saitoIsospin2021,parkFlavour2021,kwanExciton2021,liuTheories2021,brillauxAnalytical2022,songMagic2022,linSpin2022,huangObservation2022,zhangCorrelated2022,herzogReentrant2022,andreiGraphene2020,stepanovCompeting2021,panThermodynamic2022,zhangMomentum2021,zhangFermion2022,zhangSuperconductivity2022,zhangQuantum2022,chenRealization2021,linExciton2022,huangEvolution2023} and the programmable quantum simulators~\cite{verresenPrediction2021,samajdarEmergent2022} such as quantum gases coupled
to optical cavities~\cite{RitschCold2013}.  LR interactions in the forms of van der Waals, dipole-dipole and Coulomb have given rise to a plethora of correlated topological and quantum phases of matter beyond the semi-classical or mean-field type descriptions, and new theoretical paradigm that could cope with these fast emergent experimental facts are critically called for.\\

One particularly interesting direction is \added{to explore} the critical properties of \deleted{finite temperature} phase transitions with continuous symmetry breaking, outside the realm of the established Mermin-Wagner theorem. \added{For 1D LR antiferromagnetic Heisenberg chain~\cite{laflorencie2005critical} and Heisenberg ladders~\cite{AdelhardtContinuously2022} with  $1/r^{\alpha}$-form LR interactions, the phase diagram as well as the critical exponents have been addressed and it has been found that there is a upper critical value $\alpha_c$ above which there is no phase transitions for these systems. Below $\alpha_c$, the transition exists and the critical exponents are dependent on $\alpha$, as identified by both field theory analysis and numerical evidence. However, for 2D LR Heisenberg models with finite-temperature transitions,} \replaced{i}{I}t was \added{only} known that, for $D=2$ Heisenberg model with ferromagnetic LR interaction $1/r^{\alpha}$, a finite-temperature ferromagnetic phase will not exist when $\alpha \ge 4$ \added{which has been proved analytically in Ref.~\cite{BrunoAbsence2001}}, and for $\alpha \le 2$ the system is gapped due to the generalized Higgs mechanism~\cite{DiesselGeneralized2022,songDynamical2023} \added{and the finite-temperature ferromagnetic order should be allowed}. However, the situation in $\alpha \in (2,4)$ is not well understood. Although there are classical field theory predictions and renormalization group analysis on this issue~\cite{CriticalExponents1972,sak1973recursion,sakLow1977}, which state there is a Gaussian fix-point for $2<\alpha<3$ and a non-Gaussian fixed-point for $3\le \alpha <4$, a thorough numerical treatment on the 2D quantum Heisenberg model has not been performed to date. Such unbiased numerical analysis of this model is crucial not only because the field-theory scenario needs to be \replaced{impartially}{unbiasedly} examined on the realistic lattice models, but also due to the fact that the Heisenberg model is one of the most central toy models in condensed matter and statistic physics and a complete clarification of the critical properties of this model will serve as the cornerstone of further studies on LR quantum many body systems. 

\begin{figure}[htp!]
	\centering
	\includegraphics[width=\columnwidth]{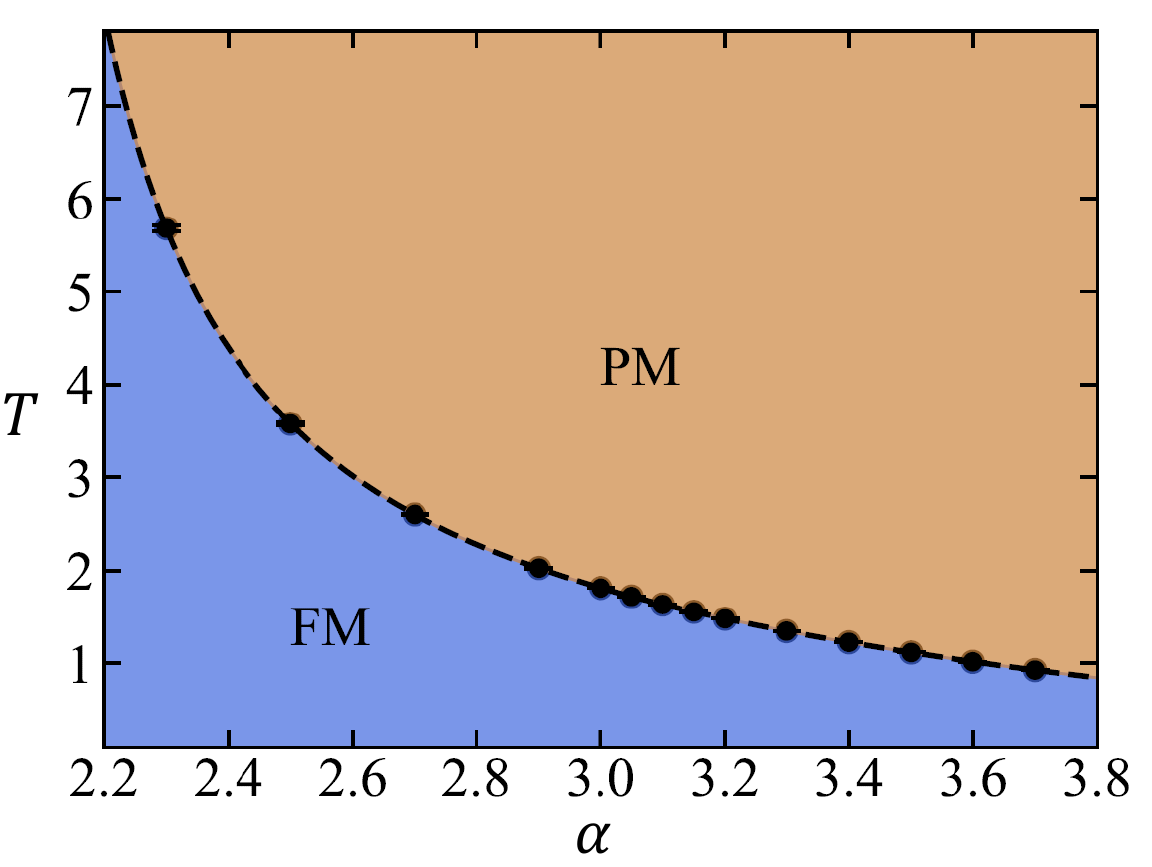}
	\caption{\textbf{Phase diagram of the 2D LR ferromagnetic Heisenberg model}. As the temperature is reduced, the system undergoes a continuous phase transition from paramagnetic phase to ferromagnetic phase in entire region of $\alpha \in (2,4)$. The black dots are the critical points determined from QMC simulations, as exemplified in Fig.~\ref{fig:fig2} and  Fig.~\ref{fig:fig3}. The standard error of the mean (SEM) is used when estimating the errors of the physical quantities.}
	\label{fig:fig1}
\end{figure}
\begin{figure}[htp!]
	\centering
	\includegraphics[width=\columnwidth]{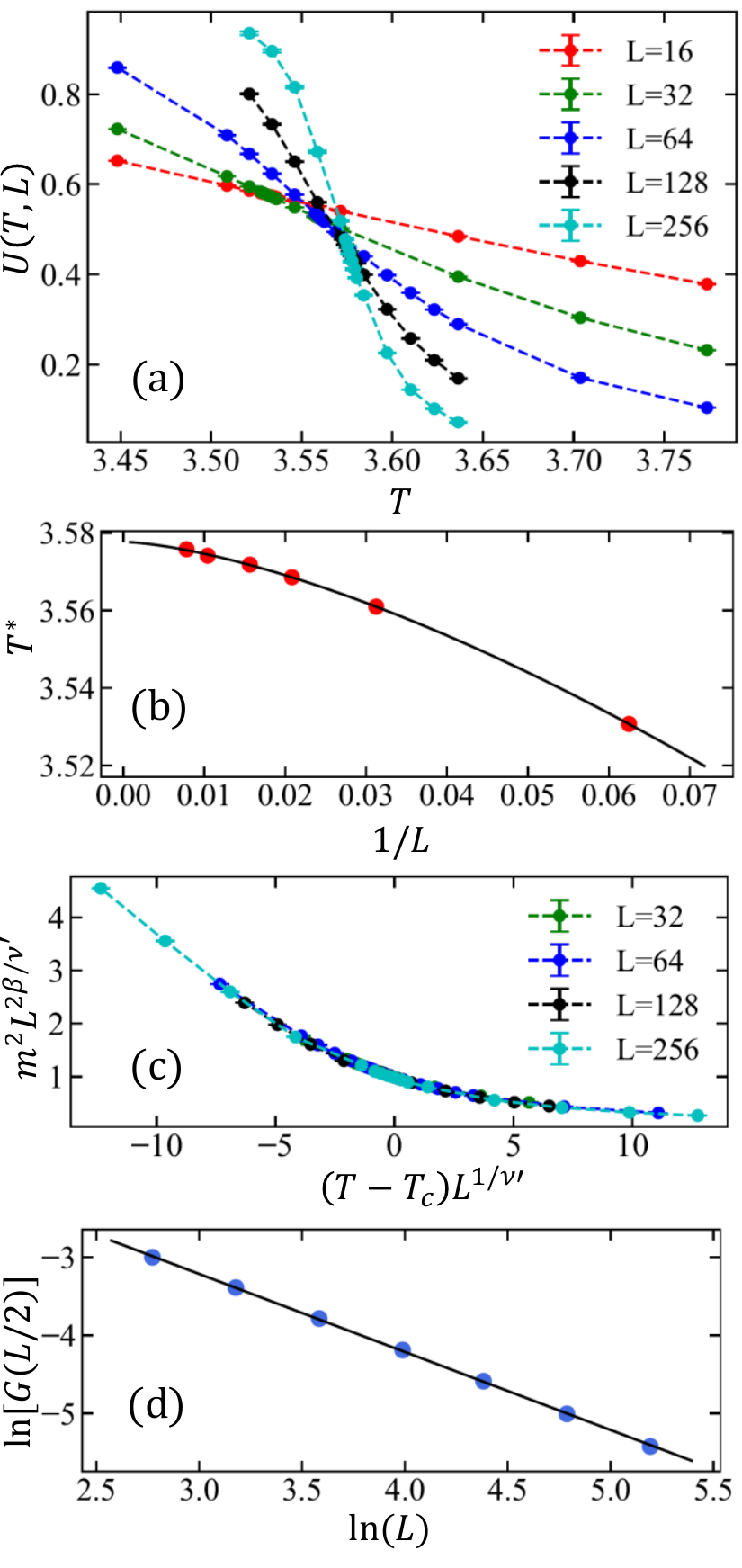}
	\caption{\textbf{The determination of the critical point and exponents at $\alpha=2.5$.} (a) Binder ratio $U(T,L)$ versus temperature $T$ for different system sizes. (b) Crossing points of Binder ratios $T^{*}(L)$ versus $1/L$. The solid line represents a fitting of the data points with Eq.~\eqref{eq:eq3}. The fitted curve is $T^{*}(L)=-2.935L^{-1.491}+3.5776$. 
		(c) Data collapse of the order parameter $\langle m^2\rangle$ near the critical point $T_c$. Notice here we replace the correlation length exponent $\nu$ with $\nu'$ as in Eq.~\eqref{eq:eq7}. 
		(d) $\ln[G(L/2)]$ versus $\ln(L)$ for different system sizes $L=16,24,36,54,80,120,180$. The data is fitted with a straight line as in Eq.~\eqref{eq:eq10} and the fitted result is $\ln[G(L/2)]=-0.999(1)\ln(L)$. The errors of $\ln[G(L/2)]$ are smaller than the symbol sizes and SEM is used when
		estimating the errors of the physical quantities.}
	\label{fig:fig2}
\end{figure}
\begin{figure}[htp!]
	\centering
	\includegraphics[width=\columnwidth]{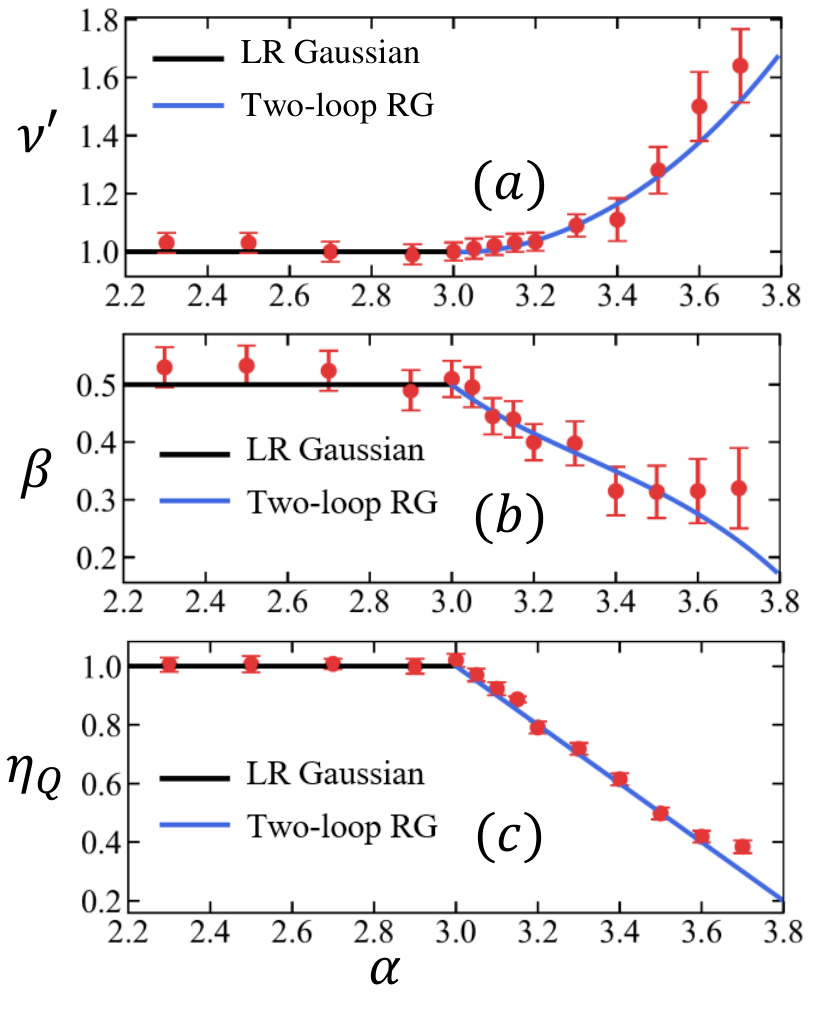}
	\caption{\textbf{Critical exponents $\nu'$, $\beta$  and $\eta_Q$} in the region of $\alpha\in[2.3,3.7]$ obtained from data collapse and from fitting to the correlation function $G(r)$.  The black and blue solid lines in (a), (b) and  (c) are the predictions of LR Gaussian theory ($\alpha<3$) and two-loop perturbative RG predictions ($3\le \alpha<4$) for Gaussian and interacting (non-Gaussian) fixed points.  \cite{CriticalExponents1972,sak1973recursion,lohmann2017critical}. SEM is used when
		estimating the errors of the physical quantities.}
	\label{fig:fig3}
\end{figure}

Here we \replaced{bridge these gaps}{address this question} by large-scale QMC simulations \added{and field theory analysis}. We find clear evidence of the  breakdown of the Mermin-Wagner theorem with finite-temperature phase transitions in $\alpha \in (2,4)$\added{, as shown in Fig.~\ref{fig:fig1}}. By performing the state-of-the-art finite-size scaling analysis, \added{as illustrated in Fig.~\ref{fig:fig2}}, we obtain the accurate critical exponents of the phase transition as a function of $\alpha$ \added{as shown in Fig.~\ref{fig:fig3}}, and demonstrate these results nicely satisfy the field-theory predictions both for $\alpha<3$ where the system is above the upper critical dimension with Gaussian fixed point and for $3\le\alpha<4$ where the system is below the upper critical dimension with non-Gaussian fixed point. Our results explicitly show the critical behaviors for $\alpha \in (2,4)$ in LR Heisenberg model at $D=2$ and will intrigue further theoretical and experimental physics and even mathematics studies of systems with LR interactions beyond the realm of the Mermin-Wagner theorem~\cite{CriticalExponents1972,sak1973recursion,sakLow1977,aizenman1988critical,abdesselam2007complete, Slade_2017,lohmann2017critical}.  \\

\noindent{\textbf{\large Results.}}\\
\noindent{\textbf{Model.}}\\
The Hamiltonian of the LR ferromagnetic Heisenberg model is 
\begin{equation}
	\H=-\sum_{i<j}J_{ij}\vec{S}_{i}\vec{S}_{j},
	\label{eq:eq1}
\end{equation}

where $J_{ij}=\frac{1}{r_{ij}^{\alpha}}$ denotes the LR coupling and $r_{ij}$ is the nearest distance between site $i$ and site $j$  under the periodic boundary condition. In order to alleviate the strong finite-size effects in systems with LR interactions \added{arising from the cut-off of LR interactions under the periodic boundary condition}, we replace $J_{ij}$ with the Ewald-corrected coupling $\tilde{J}_{ij}$~\cite{FukuiOrder2009,Flores-Sola2015} which takes the form of
\begin{equation}
	\label{Ewald-cor}
	\tilde{J}_{ij}=\sum_{m,n=-\infty}^{\infty}\frac{1}{|\vec{i}-\vec{j}+mL_{x}\vec{e_{x}}+nL_{y}\vec{e_{y}}|^{\alpha}}.
\end{equation}
This modified coupling parameter $\tilde{J}_{ij}$ counts all the possible distances between two sites under the periodic boundary condition \added{,so that the effect of cutting off the tail of LR interactions is minimized, and this trick has been shown to be very useful in the simulation of many LR systems~\cite{Flores-Sola2015,FukuiOrder2009,KoziolQuantum2021,AdelhardtContinuously2022}}. For 2D there is no closed form for Eq.~\eqref{Ewald-cor}, so we truncate the summation at $|m|_{\text{max}},|n|_{\text{max}}=1000$ for $\alpha<3$ which is large enough to have the well-converged finite-size scaling behavior, as shown in Fig.~\ref{fig:fig2}. For $\alpha\ge3$ the finite-size effects are mainly from crossovers to SR case, and we find the original coupling $J_{ij}$ is fine to obtain converged results.

When $\alpha\ge 2D$ the system reduces to the SR case where there is no spontaneously continuous symmetry breaking phase at finite-temperature. When $\alpha\le D$, the Hamiltonian is no longer extensive and there is no well-defined thermodynamic limit. Between $\alpha \in (2,4)$ we carry out the  QMC simulations \cite{SandvikQuantum1991,SandvikStochastic1999,Sandvik2003} up to the linear system size of $L=256$, as shown in Fig.~\ref{fig:fig2}, to determine the precise phase boundary as well as the critical exponents $\nu$, $\beta$ and $\eta$. Note that because of strong finite-size effects, we only compute the region of $\alpha \in [2.3,3.7]$ where our QMC simulations can obtain well-converged results. \added{The origins of finite-size effects as $\alpha$ approaches the two boundaries, $\alpha=2$ and $\alpha=4$, exhibit inherent distinctions. When $\alpha\rightarrow 2$, the finite-size effect arises from the escalating intensity of LR (long-range) interactions, which fundamentally reduces the efficiency of the Ewald-corrected scheme. Conversely, as $\alpha\rightarrow 4$, the system approaches the regime where finite-temperature phase transitions do not \replaced{exist}{occur}. Consequently, near this boundary, the convergence of data points becomes exceedingly slow to be overcome.} The results are shown in Figs.~\ref{fig:fig1} and ~\ref{fig:fig3} and will be discussed in the critical exponents section. The QMC implementation is explained in the Supplementary Note 1.

Note that when $\alpha\le D$, the Hamiltonian defined in Eq.~\eqref{eq:eq1} can actually be Kac-normalized~\cite{Eduardo2021,DefenuMetastability2021} to be extensive with the addition of a factor $\frac{N-1}{\sum_{i< j} J_{ij}}$ to the Hamiltonian. Although this is not the focus of our paper, we examine the Kac-normalized Hamiltonian \deleted{at $\alpha=1.8$} and the results are shown in \deleted{Fig.~\ref{fig:fig4} and in} the Supplementary Note 2.\\

\noindent{\textbf{Critical exponents.}}\\
Fig.~\ref{fig:fig2} shows our results at $\alpha=2.5$. We first use the crossing points  of the Binder ratios to locate the critical temperature $T_c$. The crossing points of $U(T,L)$ with $U(T,2L)$ are denoted as $T^{*}(L)$, and through fitting to Eq.~\eqref{eq:eq3} the precise value of $T_c$ can be obtained.  We then use the value of $T_c$ to perform data collapse according to Eq.~\eqref{eq:eq4} and Eq.~\eqref{eq:eq7} separately for $3\le \alpha <4$ and $\alpha<3$, to obtain the critical exponents $\nu'$ and $\beta$. To obtain the anomalous dimension $\eta_Q$, 
we measure the correlation  function $G(L/2)$ at the obtained critical temperature $T_c$ and obtain the anomalous dimension separately by fitting to Eq.~\eqref{eq:eq5} for $3\le \alpha <4$ and Eq.~\eqref{eq:eq10} for $\alpha<3$. \\

According to the conventions defined in Eq.~\eqref{redefine} and field theory results of the mean-field critical exponents in Eq.~\eqref{mf-exponent}, we can extract the expression for the three critical exponents in the Gaussian region which are $\nu'=1$, $\beta=\frac{1}{2}$ and $\eta_Q=1$. Outside the Gaussian region, we have $\eta=4-\alpha$ and $\gamma$ defined in Eq.~\eqref{gamm}, and the value of $\beta$ and $\nu$ can be obtained via solving the scaling relations between the critical exponents with $\nu=\frac{\gamma}{2-\eta}$ and $\beta=\frac{\gamma\eta}{2(2-\eta)}$.

The critical exponents we have obtained  are shown in Fig.~\ref{fig:fig3}. We find that within the region we simulated, our QMC-obtained critical exponents $\nu'(\alpha)$ , $\beta(\alpha)$, and $\eta_{Q}(\alpha)$ match nicely with the prediction of both LR Gaussian theory (for $\alpha<3$) and the two-loop perturbative RG (for $3\le\alpha<4$) \added{, although there is a sign of deviating from two-loop RG predictions when $\alpha$ approaches 4. The possible deviation might be explained by the increasing finite-size effects near the boundary or the inefficiency of two-loop perturbative RG predictions when $\alpha$ is away from $\alpha=3$. The results can be further improved by either considering higher-order RG corrections or by pushing the QMC simulations to larger system sizes. } Notably, the \added{predicted form of} anomalous dimension $\eta$ receive no corrections at any $\alpha\in(2,4)$~\cite{CriticalExponents1972}  and our results confirm this argument with $\eta$ matching with $\eta=4-\alpha$ well in the whole region.

\noindent{\textbf{\large Discussions.}}\\
\added{Our investigation reveals a finite-temperature phase transition point in the 2D LR Heisenberg model, occurring for values of $\alpha$ within the range of $\alpha\in (2,4)$, which separates the ferromagnetic phase from the paramagnetic phase. We observe that the phase transition point exhibits distinct behaviors: a Gaussian fixed point characterizes the transition for $\alpha\le3$, while a non-Gaussian fixed point emerges for $3<\alpha<4$.
Similar phenomena have been observed in various LR systems~\cite{lohmann2017critical,sakLow1977,Slade_2017,DefenuLong2021,AdelhardtContinuously2022,KoziolQuantum2021,Flores-Sola2015,Eduardo2021,laflorencie2005critical}. However, it is important to note that LR Ising-like systems differ intrinsically from LR Heisenberg-like systems. The former does not adhere to the Mermin-Wagner theorem, guaranteeing a finite-temperature transition for all $\alpha>0$, while the latter exhibits an upper critical value $\alpha_c$ beyond which the Mermin-Wagner theorem precludes the existence of phase transitions.}  \replaced{In conclusion, o}{O}ur results clearly point out the LR quantum many-body system exhibit unconventional critical properties beyond the realm of the Mermin-Wagner theorem, which are also worthwhile to pursue in \added{future} experimental\replaced{realizations}{systems} \replaced{,such as the quantum simulators.}{where the LR interactions in the forms of van der Waals, dipole-dipole and Coulomb play the dominate role. Such systems include, but not limited to, Rydberg arrays, twisted bilayer Graphene and 2D quantum Moir\'e material and quantum simulators.} \\

\noindent{\textbf{\large Methods.}}\\
\noindent{\textbf{Field theory analysis.}}\\
We review here the field theory description of the model at the thermodynamic limit dating back to Ref.~\cite{CriticalExponents1972}. The action can be written as 
\begin{equation}\label{action}
	S=\int d^Dx d^Dx' \frac{\sum_{i}\phi^{i}(x)\phi^{i}(x')}{|x-x'|^{d+\sigma}}+\lambda \int d^D x \sum_{i}\phi^{i}(x)^4,
\end{equation}
to match the lattice model, we need $\alpha=d+\sigma$. 
Under the scaling symmetry 
\begin{equation}
	x\rightarrow s x, \phi^{i} \rightarrow s^{-\Delta_{\phi}} \phi^{i},
\end{equation}
the kinetic term remains unchanged when $\Delta_{\phi}=\frac{D-\sigma}{2}$. The coupling constant of $\phi^4$ interaction, on the other hand, scale as 
\begin{equation}
	\lambda \rightarrow s^{2\alpha-3D}\lambda.
\end{equation}
When $\alpha<\frac{3D}{2}$, the coupling constant decays at larger length scale, which means the $\lambda \phi^4$ term is an irrelevant operator. The Gaussian fixed point at $\lambda=0$ is a stable fixed point. \added{Notice when $\lambda=0$, the action is in a purely quadratic form, hence named "Gaussian" fixed point.} This was established mathematically in Ref.~\cite{lohmann2017critical}. When $\alpha>\frac{3D}{2}$, the $\lambda \phi^4$ term becomes relevant, which triggers a renormalization group towards a different non-Gaussian fixed point~\cite{CriticalExponents1972}. 
One can perform standard renormalization technique to calculate the scaling dimension of various operators, by evaluating Feynman diagrams with non-conventional propagators. Such a calculation was first performed in~\cite{CriticalExponents1972}. Since the kinetic term in Eq.~\eqref{action} is no-local, which can not receive corrections from any local counter terms, the scaling dimension of $\phi$ will not be renormalized (This can be easily seen by analyzing the the Callan-Symanzik equation for the two point function $\langle\phi(x)\phi(y)\rangle$, see for example, Ref.~\cite{behan2017scaling}). Equivalently, we have $\eta=2\Delta_{\phi}-D+2$. Our numerical result clearly confirms such a theoretical prediction. For a fixed $\sigma$ in Eq.~\eqref{action}, we can define the upper critical dimension as the space-time dimension at which the $\phi^4$ term is marginal. The $\Delta_{\phi^4}=4\Delta_{\phi}=D_{uc}$ gives us 
\begin{equation}
	\label{eq:eq12}
	D_{uc}=2\sigma=2(\alpha-D).
\end{equation}
We now focus on the $D=2$ case. When $\alpha<3$, the critical behavior is controlled by the $\lambda=0$ Gaussian fixed point. The critical behavior is similar to the usual Ising model at $D>4$, due to the effect of dangerously irrelevant operators \cite{cardy1996scaling}, the critical exponents are given by  
\begin{equation}
	\label{mf-exponent}
	\nu=\frac{1}{\alpha-2},\quad \beta=\frac{1}{2},\quad \eta=4-\alpha. 
\end{equation}
For example, the $\beta=1/2$ exponent can be seen from the following argument. Deform the action \eqref{action} by a mass term $\int dx^D t \phi(x)^2$ with negative $t$ and minimize the potential, we get $\langle\phi\rangle \propto (-t/\lambda)^{\beta}$, with $\beta=1/2$. The other exponents can be calculated by similar mean field theory analysis. The critical exponent $\eta$ controls the two point function $\langle\phi(x)\phi(y)\rangle$ only at the strict thermodynamic limit. At finite sizes, the power law behaviour will be modified to \eqref{eq:eq10}, which follows from analysing the effect of dangerously irrelevant operators carefully \cite{Kenna_2014}.

When $\alpha>3$, on the other hand, the second term in Eq.~\eqref{action} becomes relevant, and renormalization group flows towards a different non-Gaussian fixed point~\cite{CriticalExponents1972}. The critical exponent $\eta$ will remain at its mean field theory value~\cite{CriticalExponents1972} as in Eq.~\eqref{mf-exponent}. 
The other exponents, on the other hand receives correction at $\mathcal{O}\left((\alpha-3)^2\right)$. The two-loop perturbation results for $\gamma$ is 
\begin{equation}
	\label{gamm}
	\frac{1}{\gamma}=1-\left(\frac{n+2}{n+8}\right) \frac{\epsilon}{\sigma}-\frac{(n+2)(7 n+20)}{(n+8)^3} Q(\sigma)\left(\frac{\epsilon}{\sigma}\right)^2+O\left(\epsilon^3\right)
\end{equation}
with $Q(\sigma)=\sigma\left[\psi(1)-2 \psi\left(\frac{1}{2} \sigma\right)+\psi(\sigma)\right]$ where $\psi(z)$ is the logarithmic derivative of the gamma function. The other critical exponents can be obtained by scaling relations between them.

When $\alpha>4$, the long\added{-}range model becomes equivalent to short\added{-}range models, due to the Mermin–Wagner theorem~\cite{merinAbsence1966,hohenbergExistence1967,halperin2019}, the system will be gapped at finite-temperature. \added{In the field-theory language, the value of $\alpha$ at which such a long-range to short-range crossover happens when the scaling dimension of $\phi$ equals to the scaling dimension of the short range model. In two dimensions, this gives $\alpha=4$ \cite{CriticalExponents1972,sak1973recursion}.}\\

\noindent{\textbf{Finite-size scaling analysis.}} \\
To identify the phase transitions and obtain the critical exponents, we compute the square magnetization $\langle m^2 \rangle$, the correlation function $G(r)$, and the Binder ratio $U(T,L)=\frac{5}{2}(1-\frac{1}{3}\frac{\langle m^4 \rangle}{\langle m^2 \rangle^2})$ in the QMC simulation. The crossing point of $U(T,L)$ with $U(T,2L)$ is denoted as $T^{*}(L)$ and it is expected to converge to the thermodynamic limit critical temperature $T_c$ following the scaling relation:
\begin{equation}
	\label{eq:eq3}
	T^{*}(L)=aL^{-b}+T_c.
\end{equation}
Given the values of $T^{*}(L)$ with sufficiently small errors and large enough system sizes $L$ , the critical point $T_c$ can be precisely located as shown in Fig.~\ref{fig:fig2}. 
To obtain the critical exponents $\nu$, $\beta$ and $\eta$, when $D\le D_{\text{uc}}$, the standard finite-size scaling behavior (FSS)~\cite{FisherCritical1972,Brezin1982} allows us to perform a data collapse near the critical points with the relation
\begin{equation}
	m^2 \sim L^{-2 \beta/\nu} \cdot f\left[L^{1/\nu}\left(T-T_c\right)\right], \quad T \sim T_c.
	\label{eq:eq4}
\end{equation}
The anomalous dimension can also be obtained by fitting to the correlation function at the critical point $T_c$
\begin{equation}
	G(r)=\langle S^{z}_{\vec{r'}}S^{z}_{\vec{r'}+\vec{r}}\rangle \sim r^{-D+2-\eta}.
	\label{eq:eq5}
\end{equation}
However, when $D>D_{\text{uc}}$, which is our case when $\alpha<3$, the system enters the mean-field region where the hyperscaling relation breaks down, famously due to the effect of dangerously irrelevant operator~\cite{Kenna_2013,Kenna_2014,Bertrand2022}. The scaling of the correlation length in this region shall follow the relation $\xi_L\sim L^{\frac{D_{\text{uc}}}{D}}$ instead of $\xi_L\sim L$~\cite{Kenna_2013,Kenna_2014,Flores-Sola2015,KoziolQuantum2021,Eduardo2021,Langheld2022,Bertrand2022}, and this leads to the modification of hyperscaling relation with
\begin{equation}
	\nu^{\prime} d=2-\alpha_H\added{,}
	\label{eq:eq6}
\end{equation}
where  $\nu'=\frac{D_{\text{uc}}}{D}\nu$ and $\alpha_H$ is the critical exponent associated with the specific heat.
For our system Eq.~\eqref{eq:eq1}, the upper critical dimension is $D_{\text{uc}}=2(\alpha-D)$, which we will explain later in the field theory analysis section. Accordingly, Eq.~\eqref{eq:eq4} also needs to be modified and the correct relation for data collapse in mean field region is\cite{Kenna_2013,Kenna_2014,Flores-Sola2015,KoziolQuantum2021,Eduardo2021}
\begin{equation}
	m^2 \sim L^{-2 \beta/\nu'} \cdot f\left[L^{1/\nu'}\left(T-T_c\right)\right], \quad T \sim T_c
	\label{eq:eq7}.
\end{equation}
The scaling of correlation function for $\alpha<3$ is also modified with
\begin{equation}
	\label{eq:eq10}
	G(r)=\langle S^{z}_{\vec{r'}}S^{z}_{\vec{r'}+\vec{r}}\rangle \sim r^{-D+2-\eta_Q},
\end{equation}
where 
\begin{equation}
	\eta_Q = \frac{D}{D_{\text{uc}}}\eta-\frac{2D}{D_{\text{uc}}}+2.
	\label{eq:eq11}
\end{equation}  
By fitting to Eq.~\eqref{eq:eq10}, the modified anomalous dimension $\eta_Q$  as well as $\eta$  can be obtained.

To unify the conventions, we define
\begin{equation}
	\label{redefine}
	\eta_Q=\left\{\begin{array}{lr}
		\frac{D}{D_{\text{uc}}}\eta-\frac{2D}{D_{\text{uc}}}+2, \text{if} \ D>D_{\text{uc}},\\
		\eta , \text{if} \ D\le D_{\text{uc}}.
	\end{array}\right.
\end{equation}
and
\begin{equation}
		\nu'=\left\{\begin{array}{lr}
		\frac{D_{\text{uc}}}{D}\nu, \text{if} \ D>D_{\text{uc}},\\
		\nu , \text{if} \ D\le D_{\text{uc}}.
	\end{array}\right.
\end{equation}
Then $\nu'$, $\beta$ and $\eta_Q$ will be obtained with the same scaling functions for both $\alpha<3$ and $3\le \alpha <4$.\\
\section*{Data availability} 
The data that support the findings of this study are available from the corresponding author upon request to the authors. \\
\section*{Code availability} 
All numerical codes in this paper are available upon request to the authors. \\
\section*{Acknowledgement}
 We thank Subir Sachdev, Fabien Alet, Fakher Assaad, Kai Sun, Michael Scherer and Lukas Janssen for valuable discussions on the related topic. JRZ thanks Mr. Tianyu Wu and Ms. Zhenzhi Qin for valuable discussions. JRZ, MHS and ZYM acknowledge the support from the Research Grants Council (RGC) of Hong Kong SAR of China (Project Nos. 17301420, 17301721, AoE/P-701/20, 17309822, HKU C7037-22G), the ANR/RGC Joint Research Scheme sponsored by RGC of Hong Kong and French National Research Agency (Project No. A\_HKU703/22), the K. C. Wong Education Foundation (Grant No. GJTD-2020-01) and the Seed Fund “Quantum-Inspired explainable-AI” at the HKU-TCL Joint Research Centre for Artificial Intelligence. The authors also acknowledge the Tianhe-II platform at the National Supercomputer Center in Guangzhou, the HPC2021 system under the Information Technology Services
and the Blackbody HPC system at the Department of Physics, University of Hong Kong for their technical support and generous allocation of CPU time.
\section*{Author contributions}
J.Z., Z.Y.M, and J.R. initiated the work. J.Z. carried out the Quantum Monte Carlo simulations. Juncheng Rong conducted the field theory analysis .All authors contributed to the analysis of the results and the preparation and revision of the draft. 

\section*{Supplementary Materials}

\vspace{\baselineskip}
\noindent {\bf Supplementary Note 1: SSE QMC  update scheme.}

The Hamiltonian of the long-range ferromagnetic Heisenberg model discussed in main text can be decomposed as diagonal and off-diagonal operators,
\begin{equation}
	\label{ferro-decom}
	\begin{split}
		H_{0, 0}&=I\\
		H_{1,a(ij)}&=J_{ij}(\frac{1}{4}+S^{z}_{i}S^{z}_{j})\\
		H_{2,a(ij)}&=\frac{J_{ij}}{2}(S^{+}_{i}S^{-}_{j}+S^{-}_{i}S^{+}_{j}),\\
	\end{split}
\end{equation}
where $H-\frac{\sum_{i< j} J_{ij}}{4}=-\sum_{a(i< j)} H_{1,a(ij)}+H_{2,a(ij)}$ and $a(ij)$ is the bond index. Take the eigenstates of $\sigma^z$ as basis, the non-zero matrix elements in Eq.~\eqref{ferro-decom} are
\begin{equation}
	\label{ferro-decom2}
	\begin{split}
		\langle \uparrow\uparrow|H_{1,a}|\uparrow\uparrow\rangle &= \langle \downarrow\downarrow|H_{1,a}|\downarrow\downarrow\rangle=\frac{J_{ij}}{2}\\
		\langle \uparrow\downarrow|H_{2,a}|\downarrow\uparrow\rangle &=\langle \downarrow\uparrow|H_{2,a}|\uparrow\downarrow\rangle=\frac{J_{ij}}{2}.\\
	\end{split}
\end{equation}
In the SSE QMC simulation, the loop update scheme is maintained the same with the ferromagnetic Heisenberg model with nearest-neighbor couplings. However, to efficiently carry out the diagonal update scheme, we choose the candidate bonds for inserting diagonal operators with an importance sampling procedure with $P_{\text{choose}}\propto J_{ij}$. The diagonal update scheme is thus
\begin{enumerate}
	\item If a diagonal operator ($H_{1,a}$) is visited, remove it with probability
	\begin{equation}
		P_{\text{remove}}=\text{min}\left(\frac{2(M-n+1)}{\beta \sum_{i< j} J_{ij}},1\right)
	\end{equation}
	\item If an identity operator ($H_{0,a}$) is visited, insert a diagonal operator according to:
	\begin{itemize}
		\item First choose a candidate bond $a$ to make the insertion with probability
		\begin{equation}
			\label{choose}
			P_{\text{choose}}=\frac{J_{ij}}{\sum_{i< j} J_{ij}}
		\end{equation}
		\item Then accept the insertion of a diagonal operator at this position with probability
		\begin{equation}
			P_{\text{accept}}=\text{min}\left(\frac{\beta \sum_{i< j} J_{ij}}{2(M-n)},1\right)
		\end{equation}
	\end{itemize}
\end{enumerate}
To generate a set of random bond index according to the probability defined in Eq.~\eqref{choose}, we use the naive Walker's method~\cite{walker1977efficient} with complexity of $O(N^2)$. Despite there is optimization of this method which reduces the complexity to $O(N)$ \cite{FukuiOrder2009}, we find for the system size we simulate the original method is sufficient and easy to implement.\\
The above procedure ensures that diagonal operators with higher matrix elements have higher probability to be inserted and compared with randomly choosing candidate bonds this strategy certainly has better efficiency. 
\begin{figure}[htp!]
	\centering
	\includegraphics[width=\columnwidth]{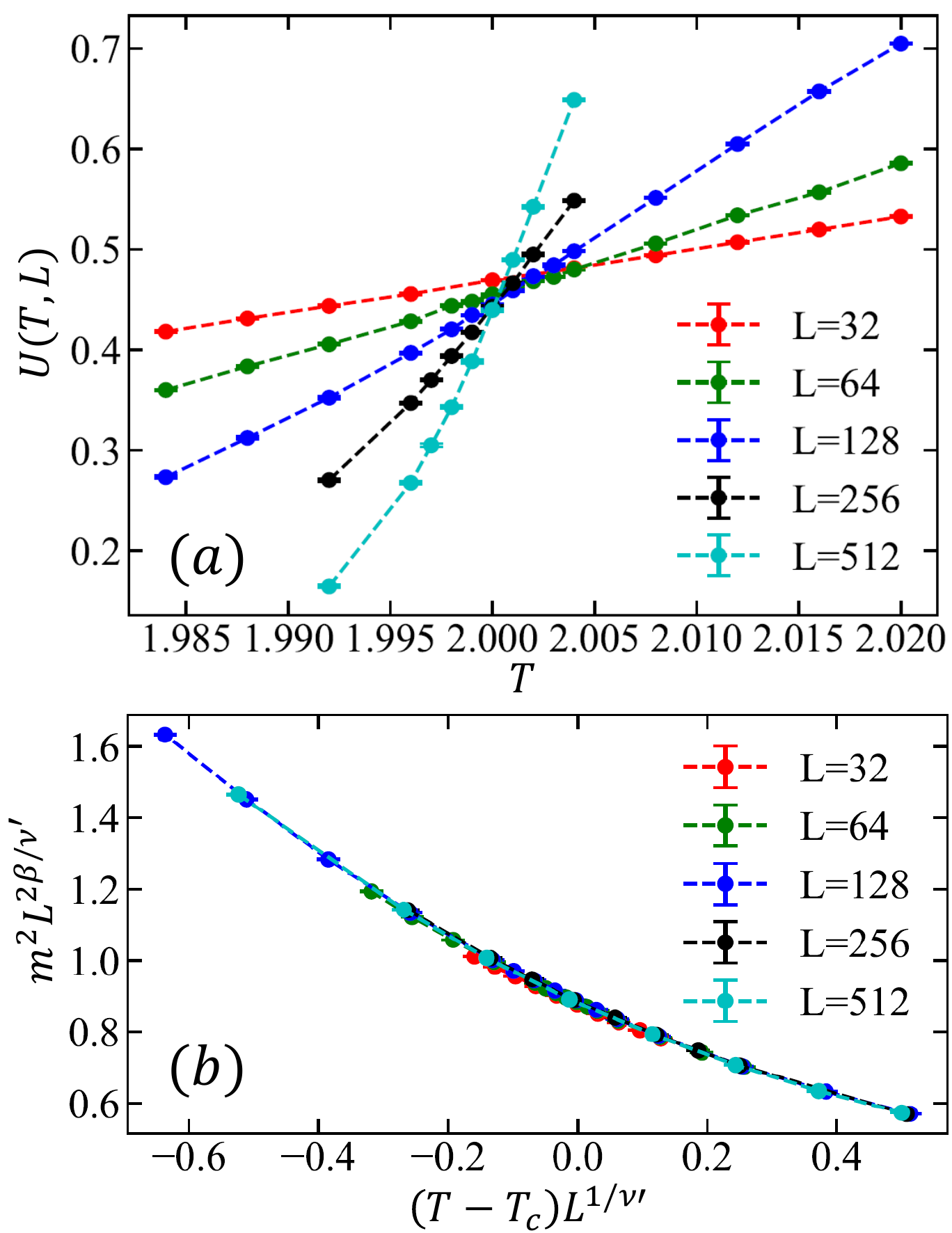}
	\caption{\textbf{The determination of the critical point and exponents at $\alpha=1.8$.} (a) Binder ratio $U(T,L)$ versus temperature $T$ for different system sizes. (b) Data collapse of the order parameter $\langle m^2\rangle$ near the critical point $T_c$. The obtained results are $\nu'=1.00(5)$ and $\beta=0.505(8)$. The standard error of the mean (SEM) is used when estimating
		the errors of the physical quantities.}
	\label{fig:figs1}
\end{figure}

\vspace{\baselineskip}
\noindent {\bf Supplementary Note 2: Kac normalization and phase diagram at $\alpha<2$.}

\begin{figure}[htp!]
	\centering
	\includegraphics[width=\columnwidth]{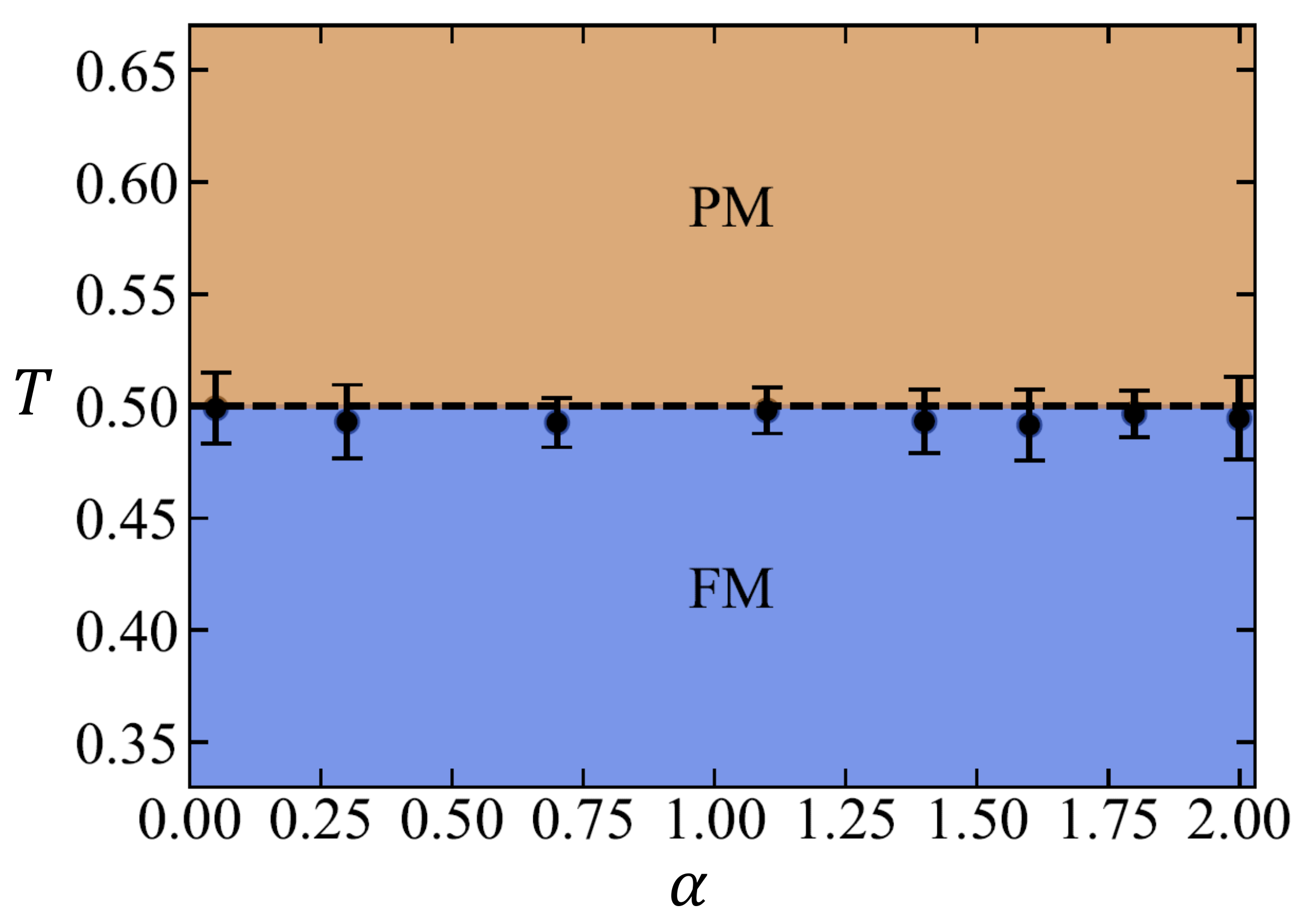}
	\caption{\textbf{Phase diagram of the 2D LR ferromagnetic Heisenberg model with Kac-normalized Hamiltonian}. The system also undergoes a continuous phase transition from FM to PM as temperature is increased. The black dots are the critical points determined from QMC simulations. The standard error of the mean (SEM) is used when estimating the errors of the physical quantities.}
	\label{fig:figs2}
\end{figure}
With a LR Hamiltonian defined as
\begin{equation}
	H=-\sum_{i<j}J_{ij}\vec{S}_{i}\vec{S}_{j}.
\end{equation}

The ground state is a fully ferromagnetic state with all spins aligned in the same direction. We have shown in the main text that the system undergoes a continuous phase transition at $\alpha \in (2,4)$. However, when $\alpha\le2$ the system is no longer extensive and it is still unclear whether such a system still hold a phase transition point. To make the system extensive, a Kac normalization factor can be added to the Hamiltonian  and the normalized Hamiltonian is
\begin{equation}
	\label{kac-H}
	H=-\frac{N-1}{\sum_{i< j} J_{ij}}\sum_{i<j}J_{ij}\vec{S}_{i}\vec{S}_{j}.
\end{equation}

In this case, the energy density  is $\langle \uparrow\cdots\uparrow\uparrow|H/N|\uparrow\uparrow\cdots \uparrow\rangle=-(N-1)/N$ which will be a constant for any value of $\alpha$. For the Hamiltonian defined in Eq.~\eqref{kac-H}, we perform the QMC simulations and find that there is a continuous phase transition for $\alpha=1.8$ and the critical exponents also satisfies the prediction of mean-field theory, as shown in Supplementary Figure~\ref{fig:figs1}. The similar phenomena has also been observed in 1D LR quantum Ising models~\cite{Eduardo2021}, where at $\alpha=0.05$ which is below the system dimension, the critical exponents are still consistent with mean-field predictions.

In addition, we also examine this system at other values of $\alpha$ and we find that the phase transition point $T_c$ remains the same for all the $\alpha \le 2$ we consider, as indicated in Supplementary Figure~\ref{fig:figs2}. There is an intuitive understanding for this finding: the Kac normalization~\cite{Eduardo2021,DefenuMetastability2021} makes the energy scale to be the same for all $\alpha \le 2$  which somehow suppresses the effect of different decaying exponent $\alpha$ and the phase transition temperature $T_c$ is thus scaled to be the same for all $\alpha$. This point should be further examined by more robust analysis.

\bibliographystyle{naturemag}
\bibliography{bibtex}

\begin{thebibliography}{10}
\expandafter\ifx\csname url\endcsname\relax
  \def\url#1{\texttt{#1}}\fi
\expandafter\ifx\csname urlprefix\endcsname\relax\def\urlprefix{URL }\fi
\providecommand{\bibinfo}[2]{#2}
\providecommand{\eprint}[2][]{\url{#2}}

\bibitem{DiesselGeneralized2022}
\bibinfo{author}{Diessel, O.~K.}, \bibinfo{author}{Diehl, S.},
  \bibinfo{author}{Defenu, N.}, \bibinfo{author}{Rosch, A.} \&
  \bibinfo{author}{Chiocchetta, A.}
\newblock \bibinfo{title}{Generalized higgs mechanism in long-range-interacting
  quantum systems}.
\newblock \emph{\bibinfo{journal}{Phys. Rev. Res.}}
  \textbf{\bibinfo{volume}{5}} (\bibinfo{year}{2023}).

\bibitem{songDynamical2023}
\bibinfo{author}{Song, M.}, \bibinfo{author}{Zhao, J.}, \bibinfo{author}{Zhou,
  C.} \& \bibinfo{author}{Meng, Z.~Y.}
\newblock \bibinfo{title}{Dynamical properties of quantum many-body systems
  with long-range interactions}.
\newblock \emph{\bibinfo{journal}{Phys. Rev. Res.}}
  \textbf{\bibinfo{volume}{5}}, \bibinfo{pages}{033046} (\bibinfo{year}{2023}).

\bibitem{CriticalExponents1972}
\bibinfo{author}{Fisher, M.~E.}, \bibinfo{author}{Ma, S.-k.} \&
  \bibinfo{author}{Nickel, B.~G.}
\newblock \bibinfo{title}{Critical exponents for long-range interactions}.
\newblock \emph{\bibinfo{journal}{Phys. Rev. Lett.}}
  \textbf{\bibinfo{volume}{29}}, \bibinfo{pages}{917--920}
  (\bibinfo{year}{1972}).

\bibitem{sak1973recursion}
\bibinfo{author}{Sak, J.}
\newblock \bibinfo{title}{Recursion relations and fixed points for ferromagnets
  with long-range interactions}.
\newblock \emph{\bibinfo{journal}{Phys. Rev. B}} \textbf{\bibinfo{volume}{8}},
  \bibinfo{pages}{281--285} (\bibinfo{year}{1973}).

\bibitem{aizenman1988critical}
\bibinfo{author}{Aizenman, M.} \& \bibinfo{author}{Fern{\'a}ndez, R.}
\newblock \bibinfo{title}{Critical exponents for long-range interactions}.
\newblock \emph{\bibinfo{journal}{Lett. Math. Phys.}}
  \textbf{\bibinfo{volume}{16}}, \bibinfo{pages}{39--49}
  (\bibinfo{year}{1988}).

\bibitem{lohmann2017critical}
\bibinfo{author}{Lohmann, M.}, \bibinfo{author}{Slade, G.} \&
  \bibinfo{author}{Wallace, B.~C.}
\newblock \bibinfo{title}{Critical two-point function for long-range o (n)
  models below the upper critical dimension}.
\newblock \emph{\bibinfo{journal}{J. Stat. Phys.}}
  \textbf{\bibinfo{volume}{169}}, \bibinfo{pages}{1132--1161}
  (\bibinfo{year}{2017}).

\bibitem{sakLow1977}
\bibinfo{author}{Sak, J.}
\newblock \bibinfo{title}{Low-temperature renormalization group for
  ferromagnets with long-range interactions}.
\newblock \emph{\bibinfo{journal}{Phys. Rev. B}} \textbf{\bibinfo{volume}{15}},
  \bibinfo{pages}{4344--4347} (\bibinfo{year}{1977}).

\bibitem{Slade_2017}
\bibinfo{author}{Slade, G.}
\newblock \bibinfo{title}{Critical exponents for long-range o(n) models below
  the upper critical dimension}.
\newblock \emph{\bibinfo{journal}{Commun. Math. Phys.}}
  \textbf{\bibinfo{volume}{358}}, \bibinfo{pages}{343--436}
  (\bibinfo{year}{2017}).

\bibitem{DefenuLong2021}
\bibinfo{author}{Defenu, N.} \emph{et~al.}
\newblock \bibinfo{title}{Long-range interacting quantum systems. {P}reprint at
  \href{https://arxiv.org/abs/2109.01063}{https://arxiv.org/abs/2109.01063}}
  (\bibinfo{year}{2023}).

\bibitem{Eduardo2021}
\bibinfo{author}{Lazo, E.~G.}, \bibinfo{author}{Heyl, M.},
  \bibinfo{author}{Dalmonte, M.} \& \bibinfo{author}{Angelone, A.}
\newblock \bibinfo{title}{{Finite-temperature critical behavior of long-range
  quantum Ising models}}.
\newblock \emph{\bibinfo{journal}{SciPost Phys.}}
  \textbf{\bibinfo{volume}{11}}, \bibinfo{pages}{076} (\bibinfo{year}{2021}).

\bibitem{BirnkammerCharacterzing2022}
\bibinfo{author}{Birnkammer, S.}, \bibinfo{author}{Bohrdt, A.},
  \bibinfo{author}{Grusdt, F.} \& \bibinfo{author}{Knap, M.}
\newblock \bibinfo{title}{Characterizing topological excitations of a
  long-range heisenberg model with trapped ions}.
\newblock \emph{\bibinfo{journal}{Phys. Rev. B}}
  \textbf{\bibinfo{volume}{105}}, \bibinfo{pages}{L241103}
  (\bibinfo{year}{2022}).

\bibitem{PeterAnomalous2012}
\bibinfo{author}{Peter, D.}, \bibinfo{author}{M\"uller, S.},
  \bibinfo{author}{Wessel, S.} \& \bibinfo{author}{B\"uchler, H.~P.}
\newblock \bibinfo{title}{Anomalous behavior of spin systems with dipolar
  interactions}.
\newblock \emph{\bibinfo{journal}{Phys. Rev. Lett.}}
  \textbf{\bibinfo{volume}{109}}, \bibinfo{pages}{025303}
  (\bibinfo{year}{2012}).

\bibitem{zhuQuantum2016}
\bibinfo{author}{Zhu, L.}, \bibinfo{author}{Hou, C.} \& \bibinfo{author}{Varma,
  C.~M.}
\newblock \bibinfo{title}{Quantum criticality in the two-dimensional
  dissipative quantum xy model}.
\newblock \emph{\bibinfo{journal}{Phys. Rev. B}} \textbf{\bibinfo{volume}{94}},
  \bibinfo{pages}{235156} (\bibinfo{year}{2016}).

\bibitem{AdelhardtContinuously2022}
\bibinfo{author}{Adelhardt, P.} \& \bibinfo{author}{Schmidt, K.~P.}
\newblock \bibinfo{title}{Continuously varying critical exponents in long-range
  quantum spin ladders}.
\newblock \emph{\bibinfo{journal}{SciPost Phys.}}
  \textbf{\bibinfo{volume}{15}}, \bibinfo{pages}{087} (\bibinfo{year}{2023}).

\bibitem{HamerThird-order1992}
\bibinfo{author}{Hamer, C.~J.}, \bibinfo{author}{Weihong, Z.} \&
  \bibinfo{author}{Arndt, P.}
\newblock \bibinfo{title}{Third-order spin-wave theory for the heisenberg
  antiferromagnet}.
\newblock \emph{\bibinfo{journal}{Phys. Rev. B}} \textbf{\bibinfo{volume}{46}},
  \bibinfo{pages}{6276--6292} (\bibinfo{year}{1992}).

\bibitem{jenkins2022breaking}
\bibinfo{author}{Jenkins, S.} \emph{et~al.}
\newblock \bibinfo{title}{Breaking through the mermin-wagner limit in 2d van
  der waals magnets}.
\newblock \emph{\bibinfo{journal}{Nat. Commun.}} \textbf{\bibinfo{volume}{13}},
  \bibinfo{pages}{6917} (\bibinfo{year}{2022}).

\bibitem{Maghrebi2017}
\bibinfo{author}{Maghrebi, M.~F.}, \bibinfo{author}{Gong, Z.-X.} \&
  \bibinfo{author}{Gorshkov, A.~V.}
\newblock \bibinfo{title}{Continuous symmetry breaking in 1d long-range
  interacting quantum systems}.
\newblock \emph{\bibinfo{journal}{Phys. Rev. Lett.}}
  \textbf{\bibinfo{volume}{119}}, \bibinfo{pages}{023001}
  (\bibinfo{year}{2017}).

\bibitem{KoziolQuantum2021}
\bibinfo{author}{Koziol, J.~A.}, \bibinfo{author}{Langheld, A.},
  \bibinfo{author}{Kapfer, S.~C.} \& \bibinfo{author}{Schmidt, K.~P.}
\newblock \bibinfo{title}{Quantum-critical properties of the long-range
  transverse-field ising model from quantum monte carlo simulations}.
\newblock \emph{\bibinfo{journal}{Phys. Rev. B}}
  \textbf{\bibinfo{volume}{103}}, \bibinfo{pages}{245135}
  (\bibinfo{year}{2021}).

\bibitem{Flores-Sola2015}
\bibinfo{author}{Flores-Sola, E.~J.}, \bibinfo{author}{Berche, B.},
  \bibinfo{author}{Kenna, R.} \& \bibinfo{author}{Weigel, M.}
\newblock \bibinfo{title}{Finite-size scaling above the upper critical
  dimension in ising models with long-range interactions}.
\newblock \emph{\bibinfo{journal}{Eur. Phys. J. B}}
  \textbf{\bibinfo{volume}{88}} (\bibinfo{year}{2015}).

\bibitem{wangOn2022}
\bibinfo{author}{{Wang}, Z.}, \bibinfo{author}{{Assaad}, F.} \&
  \bibinfo{author}{{Ulybyshev}, M.}
\newblock \bibinfo{title}{On the validity of slac fermions for the 1+1d helical
  luttinger liquid. {P}reprint at
  \href{https://arxiv.org/abs/2211.02960}{https://arxiv.org/abs/2211.02960}}
  (\bibinfo{year}{2022}).

\bibitem{WeberDissipation2022}
\bibinfo{author}{Weber, M.}, \bibinfo{author}{Luitz, D.~J.} \&
  \bibinfo{author}{Assaad, F.~F.}
\newblock \bibinfo{title}{Dissipation-induced order: The $s=1/2$ quantum spin
  chain coupled to an ohmic bath}.
\newblock \emph{\bibinfo{journal}{Phys. Rev. Lett.}}
  \textbf{\bibinfo{volume}{129}}, \bibinfo{pages}{056402}
  (\bibinfo{year}{2022}).

\bibitem{wernerQuantum2005}
\bibinfo{author}{Werner, P.}, \bibinfo{author}{Troyer, M.} \&
  \bibinfo{author}{Sachdev, S.}
\newblock \bibinfo{title}{Quantum spin chains with site dissipation}.
\newblock \emph{\bibinfo{journal}{J. Phys. Soc. Japan}}
  \textbf{\bibinfo{volume}{74}}, \bibinfo{pages}{67--70}
  (\bibinfo{year}{2005}).

\bibitem{laflorencie2005critical}
\bibinfo{author}{Laflorencie, N.}, \bibinfo{author}{Affleck, I.} \&
  \bibinfo{author}{Berciu, M.}
\newblock \bibinfo{title}{Critical phenomena and quantum phase transition in
  long range heisenberg antiferromagnetic chains}.
\newblock \emph{\bibinfo{journal}{J. Stat. Mech.: Theory Exp.}}
  \textbf{\bibinfo{volume}{2005}}, \bibinfo{pages}{P12001}
  (\bibinfo{year}{2005}).

\bibitem{samajdar2021quantum}
\bibinfo{author}{Samajdar, R.}, \bibinfo{author}{Ho, W.~W.},
  \bibinfo{author}{Pichler, H.}, \bibinfo{author}{Lukin, M.~D.} \&
  \bibinfo{author}{Sachdev, S.}
\newblock \bibinfo{title}{Quantum phases of rydberg atoms on a kagome lattice}.
\newblock \emph{\bibinfo{journal}{Proc. Natl. Acad. Sci. U.S.A.}}
  \textbf{\bibinfo{volume}{118}} (\bibinfo{year}{2021}).

\bibitem{yan2022triangular}
\bibinfo{author}{{Yan}, Z.}, \bibinfo{author}{{Samajdar}, R.},
  \bibinfo{author}{{Wang}, Y.-C.}, \bibinfo{author}{{Sachedev}, S.} \&
  \bibinfo{author}{{Meng}, Z.~Y.}
\newblock \bibinfo{title}{{Triangular lattice quantum dimer model with variable
  dimer density}}.
\newblock \emph{\bibinfo{journal}{Nat. Commun.}} \textbf{\bibinfo{volume}{13}},
  \bibinfo{pages}{5799} (\bibinfo{year}{2022}).

\bibitem{Semeghini21}
\bibinfo{author}{{Semeghini}, G.} \emph{et~al.}
\newblock \bibinfo{title}{{Probing topological spin liquids on a programmable
  quantum simulator}}.
\newblock \emph{\bibinfo{journal}{Science}} \textbf{\bibinfo{volume}{374}},
  \bibinfo{pages}{1242--1247} (\bibinfo{year}{2021}).

\bibitem{Roushan21}
\bibinfo{author}{{Satzinger}, K.~J.} \emph{et~al.}
\newblock \bibinfo{title}{{Realizing topologically ordered states on a quantum
  processor}}.
\newblock \emph{\bibinfo{journal}{Science}} \textbf{\bibinfo{volume}{374}},
  \bibinfo{pages}{1237--1241} (\bibinfo{year}{2021}).

\bibitem{yanEmergent2023}
\bibinfo{author}{Yan, Z.}, \bibinfo{author}{Wang, Y.-C.},
  \bibinfo{author}{Samajdar, R.}, \bibinfo{author}{Sachdev, S.} \&
  \bibinfo{author}{Meng, Z.~Y.}
\newblock \bibinfo{title}{Emergent glassy behavior in a kagome rydberg atom
  array}.
\newblock \emph{\bibinfo{journal}{Phys. Rev. Lett.}}
  \textbf{\bibinfo{volume}{130}}, \bibinfo{pages}{206501}
  (\bibinfo{year}{2023}).

\bibitem{tramblyLocalization2010}
\bibinfo{author}{Trambly~de Laissardière, G.}, \bibinfo{author}{Mayou, D.} \&
  \bibinfo{author}{Magaud, L.}
\newblock \bibinfo{title}{Localization of dirac electrons in rotated graphene
  bilayers}.
\newblock \emph{\bibinfo{journal}{Nano Lett.}} \textbf{\bibinfo{volume}{10}},
  \bibinfo{pages}{804 -- 808} (\bibinfo{year}{2010}).

\bibitem{rafiMoire2011}
\bibinfo{author}{Bistritzer, R.} \& \bibinfo{author}{MacDonald, A.~H.}
\newblock \bibinfo{title}{Moire bands in twisted double-layer graphene}.
\newblock \emph{\bibinfo{journal}{Proc. Natl. Acad. Sci. U.S.A.}}
  \textbf{\bibinfo{volume}{108}}, \bibinfo{pages}{12233--12237}
  (\bibinfo{year}{2011}).

\bibitem{tramblyNumerical2012}
\bibinfo{author}{Trambly~de Laissardi\`ere, G.}, \bibinfo{author}{Mayou, D.} \&
  \bibinfo{author}{Magaud, L.}
\newblock \bibinfo{title}{Numerical studies of confined states in rotated
  bilayers of graphene}.
\newblock \emph{\bibinfo{journal}{Phys. Rev. B}} \textbf{\bibinfo{volume}{86}},
  \bibinfo{pages}{125413} (\bibinfo{year}{2012}).

\bibitem{rozhkovElectronic2016}
\bibinfo{author}{Rozhkov, A.}, \bibinfo{author}{Sboychakov, A.},
  \bibinfo{author}{Rakhmanov, A.} \& \bibinfo{author}{Nori, F.}
\newblock \bibinfo{title}{Electronic properties of graphene-based bilayer
  systems}.
\newblock \emph{\bibinfo{journal}{Phys. Rep.}} \textbf{\bibinfo{volume}{648}},
  \bibinfo{pages}{1--104} (\bibinfo{year}{2016}).
\newblock \bibinfo{note}{Electronic properties of graphene-based bilayer
  systems}.

\bibitem{caoUnconventional2018}
\bibinfo{author}{Cao, Y.} \emph{et~al.}
\newblock \bibinfo{title}{Unconventional superconductivity in magic-angle
  graphene superlattices}.
\newblock \emph{\bibinfo{journal}{Nature}} \textbf{\bibinfo{volume}{556}},
  \bibinfo{pages}{43--50} (\bibinfo{year}{2018}).

\bibitem{caoCorrelated2018}
\bibinfo{author}{Cao, Y.} \emph{et~al.}
\newblock \bibinfo{title}{Correlated insulator behaviour at half-filling in
  magic-angle graphene superlattices}.
\newblock \emph{\bibinfo{journal}{Nature}} \textbf{\bibinfo{volume}{556}},
  \bibinfo{pages}{80--84} (\bibinfo{year}{2018}).

\bibitem{xieSpectroscopic2019}
\bibinfo{author}{Xie, Y.} \emph{et~al.}
\newblock \bibinfo{title}{Spectroscopic signatures of many-body correlations in
  magic-angle twisted bilayer graphene}.
\newblock \emph{\bibinfo{journal}{Nature}} \textbf{\bibinfo{volume}{572}},
  \bibinfo{pages}{101--105} (\bibinfo{year}{2019}).

\bibitem{luSuperconductors2019}
\bibinfo{author}{Lu, X.} \emph{et~al.}
\newblock \bibinfo{title}{Superconductors, orbital magnets and correlated
  states in magic-angle bilayer graphene}.
\newblock \emph{\bibinfo{journal}{Nature}} \textbf{\bibinfo{volume}{574}},
  \bibinfo{pages}{653--657} (\bibinfo{year}{2019}).

\bibitem{liaoValence2019}
\bibinfo{author}{Da~Liao, Y.}, \bibinfo{author}{Meng, Z.~Y.} \&
  \bibinfo{author}{Xu, X.~Y.}
\newblock \bibinfo{title}{Valence bond orders at charge neutrality in a
  possible two-orbital extended hubbard model for twisted bilayer graphene}.
\newblock \emph{\bibinfo{journal}{Phys. Rev. Lett.}}
  \textbf{\bibinfo{volume}{123}}, \bibinfo{pages}{157601}
  (\bibinfo{year}{2019}).

\bibitem{yankowitzTuning2019}
\bibinfo{author}{Yankowitz, M.} \emph{et~al.}
\newblock \bibinfo{title}{Tuning superconductivity in twisted bilayer
  graphene}.
\newblock \emph{\bibinfo{journal}{Science}} \textbf{\bibinfo{volume}{363}},
  \bibinfo{pages}{1059--1064} (\bibinfo{year}{2019}).

\bibitem{tomarkenElectronic2019}
\bibinfo{author}{Tomarken, S.~L.} \emph{et~al.}
\newblock \bibinfo{title}{Electronic compressibility of magic-angle graphene
  superlattices}.
\newblock \emph{\bibinfo{journal}{Phys. Rev. Lett.}}
  \textbf{\bibinfo{volume}{123}}, \bibinfo{pages}{046601}
  (\bibinfo{year}{2019}).

\bibitem{caoStrange2020}
\bibinfo{author}{Cao, Y.} \emph{et~al.}
\newblock \bibinfo{title}{Strange metal in magic-angle graphene with near
  planckian dissipation}.
\newblock \emph{\bibinfo{journal}{Phys. Rev. Lett.}}
  \textbf{\bibinfo{volume}{124}}, \bibinfo{pages}{076801}
  (\bibinfo{year}{2020}).

\bibitem{shenCorrelated2020}
\bibinfo{author}{Shen, C.} \emph{et~al.}
\newblock \bibinfo{title}{Correlated states in twisted double bilayer
  graphene}.
\newblock \emph{\bibinfo{journal}{Nat. Phys.}}  (\bibinfo{year}{2020}).

\bibitem{kevinStrongly2020}
\bibinfo{author}{Nuckolls, K.~P.} \emph{et~al.}
\newblock \bibinfo{title}{Strongly correlated chern insulators in magic-angle
  twisted bilayer graphene}.
\newblock \emph{\bibinfo{journal}{Nature}} \textbf{\bibinfo{volume}{588}},
  \bibinfo{pages}{610--615} (\bibinfo{year}{2020}).

\bibitem{chatterjeeSkyrmion2022}
\bibinfo{author}{Chatterjee, S.}, \bibinfo{author}{Ippoliti, M.} \&
  \bibinfo{author}{Zaletel, M.~P.}
\newblock \bibinfo{title}{Skyrmion superconductivity: Dmrg evidence for a
  topological route to superconductivity}.
\newblock \emph{\bibinfo{journal}{Phys. Rev. B}}
  \textbf{\bibinfo{volume}{106}}, \bibinfo{pages}{035421}
  (\bibinfo{year}{2022}).

\bibitem{khalafSoftmodes2020}
\bibinfo{author}{Khalaf, E.}, \bibinfo{author}{Bultinck, N.},
  \bibinfo{author}{Vishwanath, A.} \& \bibinfo{author}{Zaletel, M.~P.}
\newblock \bibinfo{title}{Soft modes in magic angle twisted bilayer graphene.
  {P}reprint at
  \href{https://arxiv.org/abs/2009.14827}{https://arxiv.org/abs/2009.14827}}
  (\bibinfo{year}{2020}).

\bibitem{xieNature2020}
\bibinfo{author}{Xie, M.} \& \bibinfo{author}{MacDonald, A.~H.}
\newblock \bibinfo{title}{Nature of the correlated insulator states in twisted
  bilayer graphene}.
\newblock \emph{\bibinfo{journal}{Phys. Rev. Lett.}}
  \textbf{\bibinfo{volume}{124}}, \bibinfo{pages}{097601}
  (\bibinfo{year}{2020}).

\bibitem{rozenEntropic2021}
\bibinfo{author}{Rozen, A.} \emph{et~al.}
\newblock \bibinfo{title}{Entropic evidence for a pomeranchuk effect in
  magic-angle graphene}.
\newblock \emph{\bibinfo{journal}{Nature}} \textbf{\bibinfo{volume}{592}},
  \bibinfo{pages}{214--219} (\bibinfo{year}{2021}).

\bibitem{saitoIsospin2021}
\bibinfo{author}{Saito, Y.} \emph{et~al.}
\newblock \bibinfo{title}{Isospin pomeranchuk effect in twisted bilayer
  graphene}.
\newblock \emph{\bibinfo{journal}{Nature}} \textbf{\bibinfo{volume}{592}},
  \bibinfo{pages}{220--224} (\bibinfo{year}{2021}).

\bibitem{parkFlavour2021}
\bibinfo{author}{Park, J.~M.}, \bibinfo{author}{Cao, Y.},
  \bibinfo{author}{Watanabe, K.}, \bibinfo{author}{Taniguchi, T.} \&
  \bibinfo{author}{Jarillo-Herrero, P.}
\newblock \bibinfo{title}{Flavour hund’s coupling, chern gaps and charge
  diffusivity in moiré graphene}.
\newblock \emph{\bibinfo{journal}{Nature}} \textbf{\bibinfo{volume}{592}},
  \bibinfo{pages}{43--48} (\bibinfo{year}{2021}).

\bibitem{kwanExciton2021}
\bibinfo{author}{Kwan, Y.~H.}, \bibinfo{author}{Hu, Y.},
  \bibinfo{author}{Simon, S.~H.} \& \bibinfo{author}{Parameswaran, S.~A.}
\newblock \bibinfo{title}{Exciton band topology in spontaneous quantum
  anomalous hall insulators: Applications to twisted bilayer graphene}.
\newblock \emph{\bibinfo{journal}{Phys. Rev. Lett.}}
  \textbf{\bibinfo{volume}{126}}, \bibinfo{pages}{137601}
  (\bibinfo{year}{2021}).

\bibitem{liuTheories2021}
\bibinfo{author}{Liu, J.} \& \bibinfo{author}{Dai, X.}
\newblock \bibinfo{title}{Theories for the correlated insulating states and
  quantum anomalous hall effect phenomena in twisted bilayer graphene}.
\newblock \emph{\bibinfo{journal}{Phys. Rev. B}}
  \textbf{\bibinfo{volume}{103}}, \bibinfo{pages}{035427}
  (\bibinfo{year}{2021}).

\bibitem{brillauxAnalytical2022}
\bibinfo{author}{Brillaux, E.}, \bibinfo{author}{Carpentier, D.},
  \bibinfo{author}{Fedorenko, A.~A.} \& \bibinfo{author}{Savary, L.}
\newblock \bibinfo{title}{Analytical renormalization group approach to
  competing orders at charge neutrality in twisted bilayer graphene}.
\newblock \emph{\bibinfo{journal}{Phys. Rev. Research}}
  \textbf{\bibinfo{volume}{4}}, \bibinfo{pages}{033168} (\bibinfo{year}{2022}).

\bibitem{songMagic2022}
\bibinfo{author}{Song, Z.-D.} \& \bibinfo{author}{Bernevig, B.~A.}
\newblock \bibinfo{title}{Magic-angle twisted bilayer graphene as a topological
  heavy fermion problem}.
\newblock \emph{\bibinfo{journal}{Phys. Rev. Lett.}}
  \textbf{\bibinfo{volume}{129}}, \bibinfo{pages}{047601}
  (\bibinfo{year}{2022}).

\bibitem{linSpin2022}
\bibinfo{author}{Lin, J.-X.} \emph{et~al.}
\newblock \bibinfo{title}{Spin-orbit--driven ferromagnetism at half moir{\'e}
  filling in magic-angle twisted bilayer graphene}.
\newblock \emph{\bibinfo{journal}{Science}} \textbf{\bibinfo{volume}{375}},
  \bibinfo{pages}{437--441} (\bibinfo{year}{2022}).

\bibitem{huangObservation2022}
\bibinfo{author}{Huang, T.} \emph{et~al.}
\newblock \bibinfo{title}{Observation of chiral and slow plasmons in twisted
  bilayer graphene}.
\newblock \emph{\bibinfo{journal}{Nature}} \textbf{\bibinfo{volume}{605}},
  \bibinfo{pages}{63--68} (\bibinfo{year}{2022}).

\bibitem{zhangCorrelated2022}
\bibinfo{author}{Zhang, S.}, \bibinfo{author}{Lu, X.} \& \bibinfo{author}{Liu,
  J.}
\newblock \bibinfo{title}{Correlated insulators, density wave states, and their
  nonlinear optical response in magic-angle twisted bilayer graphene}.
\newblock \emph{\bibinfo{journal}{Phys. Rev. Lett.}}
  \textbf{\bibinfo{volume}{128}}, \bibinfo{pages}{247402}
  (\bibinfo{year}{2022}).

\bibitem{herzogReentrant2022}
\bibinfo{author}{Herzog-Arbeitman, J.}, \bibinfo{author}{Chew, A.},
  \bibinfo{author}{Efetov, D.~K.} \& \bibinfo{author}{Bernevig, B.~A.}
\newblock \bibinfo{title}{Reentrant correlated insulators in twisted bilayer
  graphene at 25 t ($2\ensuremath{\pi}$ flux)}.
\newblock \emph{\bibinfo{journal}{Phys. Rev. Lett.}}
  \textbf{\bibinfo{volume}{129}}, \bibinfo{pages}{076401}
  (\bibinfo{year}{2022}).

\bibitem{andreiGraphene2020}
\bibinfo{author}{Andrei, E.~Y.} \& \bibinfo{author}{MacDonald, A.~H.}
\newblock \bibinfo{title}{Graphene bilayers with a twist}.
\newblock \emph{\bibinfo{journal}{Nat. Mater.}} \textbf{\bibinfo{volume}{19}},
  \bibinfo{pages}{1265 -- 1275} (\bibinfo{year}{2020}).

\bibitem{stepanovCompeting2021}
\bibinfo{author}{Stepanov, P.} \emph{et~al.}
\newblock \bibinfo{title}{Competing zero-field chern insulators in
  superconducting twisted bilayer graphene}.
\newblock \emph{\bibinfo{journal}{Phys. Rev. Lett.}}
  \textbf{\bibinfo{volume}{127}}, \bibinfo{pages}{197701}
  (\bibinfo{year}{2021}).

\bibitem{panThermodynamic2022}
\bibinfo{author}{Pan, G.} \emph{et~al.}
\newblock \bibinfo{title}{Thermodynamic characteristic for a correlated
  flat-band system with a quantum anomalous hall ground state}.
\newblock \emph{\bibinfo{journal}{Phys. Rev. Lett.}}
  \textbf{\bibinfo{volume}{130}}, \bibinfo{pages}{016401}
  (\bibinfo{year}{2023}).

\bibitem{zhangMomentum2021}
\bibinfo{author}{Zhang, X.}, \bibinfo{author}{Pan, G.}, \bibinfo{author}{Zhang,
  Y.}, \bibinfo{author}{Kang, J.} \& \bibinfo{author}{Meng, Z.~Y.}
\newblock \bibinfo{title}{Momentum space quantum monte carlo on twisted bilayer
  graphene}.
\newblock \emph{\bibinfo{journal}{Chin. Phys. Lett.}}
  \textbf{\bibinfo{volume}{38}}, \bibinfo{pages}{077305}
  (\bibinfo{year}{2021}).

\bibitem{zhangFermion2022}
\bibinfo{author}{Zhang, X.}, \bibinfo{author}{Pan, G.}, \bibinfo{author}{Xu,
  X.~Y.} \& \bibinfo{author}{Meng, Z.~Y.}
\newblock \bibinfo{title}{Fermion sign bounds theory in quantum monte carlo
  simulation}.
\newblock \emph{\bibinfo{journal}{Phys. Rev. B}}
  \textbf{\bibinfo{volume}{106}}, \bibinfo{pages}{035121}
  (\bibinfo{year}{2022}).

\bibitem{zhangSuperconductivity2022}
\bibinfo{author}{Zhang, X.}, \bibinfo{author}{Sun, K.}, \bibinfo{author}{Li,
  H.}, \bibinfo{author}{Pan, G.} \& \bibinfo{author}{Meng, Z.~Y.}
\newblock \bibinfo{title}{Superconductivity and bosonic fluid emerging from
  moir\'e flat bands}.
\newblock \emph{\bibinfo{journal}{Phys. Rev. B}}
  \textbf{\bibinfo{volume}{106}}, \bibinfo{pages}{184517}
  (\bibinfo{year}{2022}).

\bibitem{zhangQuantum2022}
\bibinfo{author}{{Zhang}, X.} \emph{et~al.}
\newblock \bibinfo{title}{Quantum monte carlo sign bounds, topological mott
  insulator and thermodynamic transitions in twisted bilayer graphene model.
  {P}reprint at
  \href{https://arxiv.org/abs/2210.11733}{https://arxiv.org/abs/2210.11733}}
  (\bibinfo{year}{2022}).

\bibitem{chenRealization2021}
\bibinfo{author}{Chen, B.-B.} \emph{et~al.}
\newblock \bibinfo{title}{Realization of topological mott insulator in a
  twisted bilayer graphene lattice model}.
\newblock \emph{\bibinfo{journal}{Nat. Commun.}} \textbf{\bibinfo{volume}{12}},
  \bibinfo{pages}{5480} (\bibinfo{year}{2021}).

\bibitem{linExciton2022}
\bibinfo{author}{Lin, X.}, \bibinfo{author}{Chen, B.-B.}, \bibinfo{author}{Li,
  W.}, \bibinfo{author}{Meng, Z.~Y.} \& \bibinfo{author}{Shi, T.}
\newblock \bibinfo{title}{Exciton proliferation and fate of the topological
  mott insulator in a twisted bilayer graphene lattice model}.
\newblock \emph{\bibinfo{journal}{Phys. Rev. Lett.}}
  \textbf{\bibinfo{volume}{128}}, \bibinfo{pages}{157201}
  (\bibinfo{year}{2022}).

\bibitem{huangEvolution2023}
\bibinfo{author}{{Huang}, C.} \emph{et~al.}
\newblock \bibinfo{title}{Evolution from quantum anomalous hall insulator to
  heavy-fermion semimetal in twisted bilayer graphene. {P}reprint at
  \href{https://arxiv.org/abs/2304.14064}{https://arxiv.org/abs/2304.14064}}
  (\bibinfo{year}{2023}).

\bibitem{verresenPrediction2021}
\bibinfo{author}{Verresen, R.}, \bibinfo{author}{Lukin, M.~D.} \&
  \bibinfo{author}{Vishwanath, A.}
\newblock \bibinfo{title}{Prediction of toric code topological order from
  rydberg blockade}.
\newblock \emph{\bibinfo{journal}{Phys. Rev. X}} \textbf{\bibinfo{volume}{11}},
  \bibinfo{pages}{031005} (\bibinfo{year}{2021}).

\bibitem{samajdarEmergent2022}
\bibinfo{author}{{Samajdar}, R.}, \bibinfo{author}{{Joshi}, D.~G.},
  \bibinfo{author}{{Teng}, Y.} \& \bibinfo{author}{{Sachdev}, S.}
\newblock \bibinfo{title}{Emergent $\mathbb{Z}_2$ gauge theories and
  topological excitations in rydberg atom arrays. {P}reprint at
  \href{https://arxiv.org/abs/2204.00632}{https://arxiv.org/abs/2204.00632}}
  (\bibinfo{year}{2022}).

\bibitem{RitschCold2013}
\bibinfo{author}{Ritsch, H.}, \bibinfo{author}{Domokos, P.},
  \bibinfo{author}{Brennecke, F.} \& \bibinfo{author}{Esslinger, T.}
\newblock \bibinfo{title}{Cold atoms in cavity-generated dynamical optical
  potentials}.
\newblock \emph{\bibinfo{journal}{Rev. Mod. Phys.}}
  \textbf{\bibinfo{volume}{85}}, \bibinfo{pages}{553--601}
  (\bibinfo{year}{2013}).

\bibitem{BrunoAbsence2001}
\bibinfo{author}{Bruno, P.}
\newblock \bibinfo{title}{Absence of spontaneous magnetic order at nonzero
  temperature in one- and two-dimensional heisenberg and $\mathit{XY}$ systems
  with long-range interactions}.
\newblock \emph{\bibinfo{journal}{Phys. Rev. Lett.}}
  \textbf{\bibinfo{volume}{87}}, \bibinfo{pages}{137203}
  (\bibinfo{year}{2001}).

\bibitem{abdesselam2007complete}
\bibinfo{author}{Abdesselam, A.}
\newblock \bibinfo{title}{A complete renormalization group trajectory between
  two fixed points}.
\newblock \emph{\bibinfo{journal}{Commun. Math. Phys.}}
  \textbf{\bibinfo{volume}{276}}, \bibinfo{pages}{727--772}
  (\bibinfo{year}{2007}).

\bibitem{FukuiOrder2009}
\bibinfo{author}{Fukui, K.} \& \bibinfo{author}{Todo, S.}
\newblock \bibinfo{title}{Order-n cluster monte carlo method for spin systems
  with long-range interactions}.
\newblock \emph{\bibinfo{journal}{J. Comput. Phys.}}
  \textbf{\bibinfo{volume}{228}}, \bibinfo{pages}{2629--2642}
  (\bibinfo{year}{2009}).

\bibitem{SandvikQuantum1991}
\bibinfo{author}{Sandvik, A.~W.} \& \bibinfo{author}{Kurkij\"arvi, J.}
\newblock \bibinfo{title}{Quantum {M}onte {C}arlo simulation method for spin
  systems}.
\newblock \emph{\bibinfo{journal}{Phys. Rev. B}} \textbf{\bibinfo{volume}{43}},
  \bibinfo{pages}{5950--5961} (\bibinfo{year}{1991}).

\bibitem{SandvikStochastic1999}
\bibinfo{author}{Sandvik, A.~W.}
\newblock \bibinfo{title}{Stochastic series expansion method with operator-loop
  update}.
\newblock \emph{\bibinfo{journal}{Phys. Rev. B}} \textbf{\bibinfo{volume}{59}},
  \bibinfo{pages}{R14157--R14160} (\bibinfo{year}{1999}).

\bibitem{Sandvik2003}
\bibinfo{author}{Sandvik, A.~W.}
\newblock \bibinfo{title}{Stochastic series expansion method for quantum ising
  models with arbitrary interactions}.
\newblock \emph{\bibinfo{journal}{Phys. Rev. E}} \textbf{\bibinfo{volume}{68}},
  \bibinfo{pages}{056701} (\bibinfo{year}{2003}).

\bibitem{DefenuMetastability2021}
\bibinfo{author}{Defenu, N.}
\newblock \bibinfo{title}{Metastability and discrete spectrum of long-range
  systems}.
\newblock \emph{\bibinfo{journal}{Proc. Natl. Acad. Sci. U.S.A.}}
  \textbf{\bibinfo{volume}{118}} (\bibinfo{year}{2021}).

\bibitem{behan2017scaling}
\bibinfo{author}{Behan, C.}, \bibinfo{author}{Rastelli, L.},
  \bibinfo{author}{Rychkov, S.} \& \bibinfo{author}{Zan, B.}
\newblock \bibinfo{title}{A scaling theory for the long-range to short-range
  crossover and an infrared duality}.
\newblock \emph{\bibinfo{journal}{J. Phys. A Math. Theor.}}
  \textbf{\bibinfo{volume}{50}}, \bibinfo{pages}{354002}
  (\bibinfo{year}{2017}).

\bibitem{cardy1996scaling}
\bibinfo{author}{Cardy, J.}
\newblock \emph{\bibinfo{title}{Scaling and renormalization in statistical
  physics}}, vol.~\bibinfo{volume}{5} (\bibinfo{publisher}{Cambridge university
  press}, \bibinfo{year}{1996}).

\bibitem{Kenna_2014}
\bibinfo{author}{Kenna, R.} \& \bibinfo{author}{Berche, B.}
\newblock \bibinfo{title}{Fisher's scaling relation above the upper critical
  dimension}.
\newblock \emph{\bibinfo{journal}{EPL}} \textbf{\bibinfo{volume}{105}},
  \bibinfo{pages}{26005} (\bibinfo{year}{2014}).

\bibitem{merinAbsence1966}
\bibinfo{author}{Mermin, N.~D.} \& \bibinfo{author}{Wagner, H.}
\newblock \bibinfo{title}{Absence of ferromagnetism or antiferromagnetism in
  one- or two-dimensional isotropic heisenberg models}.
\newblock \emph{\bibinfo{journal}{Phys. Rev. Lett.}}
  \textbf{\bibinfo{volume}{17}}, \bibinfo{pages}{1133--1136}
  (\bibinfo{year}{1966}).

\bibitem{hohenbergExistence1967}
\bibinfo{author}{Hohenberg, P.~C.}
\newblock \bibinfo{title}{Existence of long-range order in one and two
  dimensions}.
\newblock \emph{\bibinfo{journal}{Phys. Rev.}} \textbf{\bibinfo{volume}{158}},
  \bibinfo{pages}{383--386} (\bibinfo{year}{1967}).

\bibitem{halperin2019}
\bibinfo{author}{Halperin, B.~I.}
\newblock \bibinfo{title}{On the hohenberg–mermin–wagner theorem and its
  limitations}.
\newblock \emph{\bibinfo{journal}{J. Stat. Phys.}}
  \textbf{\bibinfo{volume}{175}}, \bibinfo{pages}{521 -- 529}
  (\bibinfo{year}{2019}).

\bibitem{FisherCritical1972}
\bibinfo{author}{Fisher, M.~E.}, \bibinfo{author}{Ma, S.-k.} \&
  \bibinfo{author}{Nickel, B.~G.}
\newblock \bibinfo{title}{Critical exponents for long-range interactions}.
\newblock \emph{\bibinfo{journal}{Phys. Rev. Lett.}}
  \textbf{\bibinfo{volume}{29}}, \bibinfo{pages}{917--920}
  (\bibinfo{year}{1972}).

\bibitem{Brezin1982}
\bibinfo{author}{Brézin, E.}
\newblock \bibinfo{title}{An investigation of finite size scaling}.
\newblock \emph{\bibinfo{journal}{J. Phys. France}}
  \textbf{\bibinfo{volume}{43}}, \bibinfo{pages}{15--22}
  (\bibinfo{year}{1982}).

\bibitem{Kenna_2013}
\bibinfo{author}{Kenna} \& \bibinfo{author}{Berche}.
\newblock \bibinfo{title}{A new critical exponent 'coppa' and its logarithmic
  counterpart 'hat coppa'}.
\newblock \emph{\bibinfo{journal}{Condens. Matter Phys.}}
  \textbf{\bibinfo{volume}{16}}, \bibinfo{pages}{23601} (\bibinfo{year}{2013}).

\bibitem{Bertrand2022}
\bibinfo{author}{Berche, B.}, \bibinfo{author}{Ellis, T.},
  \bibinfo{author}{Holovatch, Y.} \& \bibinfo{author}{Kenna, R.}
\newblock \bibinfo{title}{{Phase transitions above the upper critical
  dimension}}.
\newblock \emph{\bibinfo{journal}{SciPost Phys. Lect. Notes}}
  \bibinfo{pages}{60} (\bibinfo{year}{2022}).

\bibitem{Langheld2022}
\bibinfo{author}{Langheld, A.}, \bibinfo{author}{Koziol, J.~A.},
  \bibinfo{author}{Adelhardt, P.}, \bibinfo{author}{Kapfer, S.~C.} \&
  \bibinfo{author}{Schmidt, K.~P.}
\newblock \bibinfo{title}{{Scaling at quantum phase transitions above the upper
  critical dimension}}.
\newblock \emph{\bibinfo{journal}{SciPost Phys.}}
  \textbf{\bibinfo{volume}{13}}, \bibinfo{pages}{088} (\bibinfo{year}{2022}).

\bibitem{walker1977efficient}
\bibinfo{author}{Walker, A.~J.}
\newblock \bibinfo{title}{An efficient method for generating discrete random
  variables with general distributions}.
\newblock \emph{\bibinfo{journal}{ACM Trans. Math. Softw.}}
  \textbf{\bibinfo{volume}{3}}, \bibinfo{pages}{253--256}
  (\bibinfo{year}{1977}).

\end{thebibliography}

\end{document}